\documentclass[aps,prd,preprintnumbers,nofootinbib]{revtex4}
\usepackage{bm}
\usepackage{latexsym}
\usepackage{dcolumn}
\usepackage{amsmath,amsfonts,amssymb}
\usepackage{graphicx,epsfig}
\usepackage{psfrag}
\usepackage{color}
\usepackage{soul}
\usepackage{ulem}
\usepackage{amsthm}

\begin{document}
%
\title{Dimensional reduction via a novel Higgs mechanism}

\author{Saurya Das $^1$
}
\email{saurya.das@uleth.ca}

\author{Mir Faizal $^{1,2}$
}
\email{mir.faizal@uleth.ca}
\affiliation{
$
^1$Theoretical Physics Group and Quantum Alberta,
Department of Physics and Astronomy, University of Lethbridge, \\
4401 University Drive, Lethbridge, Alberta, T1K 3M4, Canada  \\
$^{2}$Irving K. Barber School of Arts and Sciences, \\ University of British Columbia - Okanagan,
\\ Kelowna, British Columbia V1V 1V7, Canada}

\begin{abstract}
Many theories of quantum gravity live in higher dimensions,
and their reduction to four dimensions
via mechanisms such as Kaluza-Klein compactification or brane world models
have associated problems.
We propose a novel mechanism of dimensional reduction via
spontaneous symmetry breaking of a higher dimensional local Lorentz group to one in lower dimensions.
Working in the gauge theory formulation of gravity, we couple
a Higgs field to spin connections,
include a potential for the field, and show that for a
suitable choice of Higgs vacuum,
the local Lorentz symmetry of the action gets spontaneously
reduced to one in a lower dimension.
Thus effectively the dimension of spacetime gets reduced by one.
This provides a viable mechanism for the dimensional reduction, and may have applications in theories of quantum gravity.
\end{abstract}

\maketitle

Quantization of gravity and its unification with the other three forces of nature remain as one of the most important problems of theoretical physics.
In many approaches to quantum gravity, dimensional reduction plays an important role. For example in string theory or $M$-theory, one starts with ten or eleven spacetime dimensions respectively, and reduces it to four dimensions via Kaluza-Klein compactification \cite{st,ss}.
This gives rise to a tower of Planck mass states which are yet unobservable.
Alternatives such as domain walls and brane world models postulate that the observable universe is constrained to live on a
four-dimensional brane, while gravity propagates in five or higher dimensions \cite{ruba,add,rs}.
These theories predict mini black holes at accessible energy scales which have not been detected, and are also severely constrained by other observational data \cite{brane2,brane3}.
In other words, dimensions greater than four play a crucial role in these theories, and bringing it down to four gives rise to new problems.

In this paper, we show that there is yet another viable method of dimensional reduction.
This is via spontaneous symmetry breaking (SSB).
One starts with a theory which has a local Lorentz symmetry
corresponding to a given spacetime dimension. Due to SSB,
some of the gravitational degrees of freedom (d.o.f.)
become massive and short ranged, and the residual
Lorentz group corresponding to the remaining long-ranged
d.o.f. is one in a lower dimension.
In other words, the gravity action (or for any other interaction)
with the above symmetry effectively sees the lower dimension.
Although gravitational Higgs mechanism has been studied in the past
\cite{earlier}-\cite{ealier1},
to our knowledge this is the first time that it is being done 
with the aim of dimensional
reduction.
A precursor to this work appeared in \cite{dfv}.
In this formalism, gravity closely resembles a gauge theory.
%
We start with such an action with a Higgs field
in $(d+1)-$spacetime dimensions:
\begin{equation}
 S =
 \int d^{d+1} x~e
\left[ f(R)
+ (D_\mu \Phi)^\dagger (D^\mu  \Phi) - V(\Phi^\dagger \Phi)
\right]~.
\label{act1}
\end{equation}
In the above, the $f(R)$ incorporates
higher order gravity theories.
Note that this (the `kinetic') part of the action
plays a spectator role in SSB.
We also work in the first order formalism of gravity in terms of the spin connections
$\omega_\mu^{ab}$, such that the curvature,
$ R^{ab}_{\mu\nu} = \partial_\mu \omega_\nu^{ab} - \partial_\nu \omega_\mu^{ab} +
 \omega^{ac}_\mu \omega^{b}_{c\nu} -  \omega^{ac}_\nu \omega^{b}_{c\mu}$.
The vielbein (vierbein for $d+1=4$) $e^\mu_a$ is defined by
$g_{\mu\nu} e^\mu_a e^\nu_b = \eta_{ab}$, with $e=\sqrt{|g|}~.$
In the above,
$\mu,\nu,$ are spacetime indices and
$a,b,$ are tangent space indices.
Furthermore, the covariant derivative is given by
$D_\mu = \partial_\mu  + \frac{i}{2} \omega^{ab}_\mu \Sigma_{ab} \equiv
 \partial_\mu + \frac{i}{2}\omega_\mu$,
where $\Sigma_{ab} = -i [\gamma_a, \gamma_b]/4$,
$ [D_\mu, D_\nu] = A^c_{\mu \nu}D_c + \frac{i}{2}R^{ab}_{\mu\nu} \Sigma_{ab}$
(we then set the torsion $A^c_{\mu \nu}$ to zero).
Varying the action with respect to the spin-connections yield the relation
$\omega_\mu^{ab}= - e^{\nu a} \nabla_\mu e^b_\nu$,
while varying with respect to the vielbeins
give rise to the Einstein equations \cite{ramond}. Here
$\Phi$ is a real field which transforms under the vector representation of $SO(d,1)$, which is the gauge symmetry group of action (\ref{act1}).
Under a gauge transformation
$U(x)=\exp(i\Lambda^{ab} (x) \Sigma_{ab}/2)$, where  $\Lambda^{ab}(x)$ is the Lorentz gauge-valued spacetime-dependent parameter,
$ \omega_\mu^{ab} \to [ U  \omega_\mu U^{-1} -  (\partial_\mu U)  U^{-1}  ]^{ab}$,
$R^{ab}_{\mu\nu} \to  [U]^a_c R_{\mu\nu}^{cd} [U^{-1}]^{b  }_{d }$
and
$\Phi \rightarrow U\Phi$.

For the potential, we choose a
non-negative function:
$V(\phi) = \frac{m^2}{2\phi_0^2}
\left[\Phi^\dagger \Phi - \phi_0^2 \right]^2
\geq 0$, where $m$ and $\phi_0$ are constants.
This is
minimized by the following Higgs field
configuration, chosen
using the aforementioned gauge freedom in
the so-called unitary gauge \cite{cottingham,chengli}:
\begin{eqnarray}
\Phi = \begin{pmatrix}
0 \\
0 \\
\cdots \\
\phi_{0} + \frac{h}{\sqrt{2}}
\end{pmatrix} ~.
\label{mat4}
\end{eqnarray}
In the above, $h$ signifies quantum fluctuations over the vacuum $\phi_0$.
Substituting (\ref{mat4}) in Eq.(\ref{act1})
yields the
spontaneously broken matter part of the
Lagrangian
\begin{eqnarray}
\mathcal{L}
&& = (D_\mu \Phi)^\dagger (D^\mu  \Phi)
- V(\Phi^\dagger \Phi) \nonumber \\
&& =
\frac{\phi_{0}^2}{4}
\left[
\omega_\mu^{0d}~\omega^{0d \mu}
+ \omega_\mu^{1d}~\omega^{1d \mu}
+ \dots
+ \omega_\mu^{d-1,d}~\omega^{d-1,d \mu}
\right] \nonumber \\
&& - m^2 h^2
+ \frac{1}{2}~\partial_\mu h \partial^\mu h
\label{scalar4} \\
&&
+ \left[ \frac{h^2}{8} + \frac{h\phi_{0}}{2\sqrt{2}}
\right]
\left[
\omega_\mu^{0d}~\omega^{0d \mu}
+ \omega_\mu^{1d}~\omega^{1d \mu}
+ \dots
+ \omega_\mu^{d-1,d}~\omega^{d-1,d \mu}
\right]
- \frac{m^2}{\phi_0^2}
\left[ \frac{1}{\sqrt{2}}~\phi_0 h^3 + \frac{1}{8} h^4 \right]~.\nonumber
\end{eqnarray}
The first line in Eq.(\ref{scalar4}) shows the the $d$ spin-connections $\omega_{\mu}^{ad}$, with $a=0,\dots,d-1$,
have each acquired a mass $M_\omega= \phi_0/\sqrt{2}$. They therefore represent a sector of the full
gravitational field which becomes short ranged
as a result of SSB.
The Higgs field acquires a mass $m$,
as seen from the second line.
The remaining terms in the third line
signify interactions between the Higgs field and
spin connections.
The other $d(d-1)/2$ spin connections
remain massless, accounting for the observed long-ranged
nature of gravity in one lower dimension.
The symmetry of the theory gets
spontaneously reduced from
$SO(d,1) \rightarrow SO(d-1,1)$.
The d.o.f. counting before and after SSB
match as well.
One starts with $(d-1)$ d.o.f. for each of the $d(d+1)/2$ massless spin connection and one for each of the $(d+1)$ Higgs components,
resulting in
$d(d+1)(d-1)/2+(d+1)=(d^3+d)/2 + 1$ d.o.f.
After SSB, for the $d$ massive spin connections
(each with $d$ d.o.f.), $d(d-1)/2$
massless spin connections (each with
$(d-1)$ d.o.f.)
and one residual massless Higgs field,
giving rise to
$d\times d + d(d-1)/2\times (d-1) +1 =(d^3+d)/2 + 1$
d.o.f. as well.

It is natural to equate the SSB scale to the
Planck scale in $d$-spacetime dimensions
(gauge group $SO(d-1,1)$), and so the mass acquired by the spin connections is of the order of Planck mass $M_\omega=M_{Pl}^{(d)}$.
This implies that they cannot be accessed by low-energy phenomena,
and the dynamics in spacetime is effectively described by a lower, $d$-dimensional theory
(gauge group $SO(d-1,1)$).
In other words, SSB has caused an effective dimensional reduction from $(d+1) \to d $ spacetime dimensions.
For example, one can start with a $5$-dimensional action with gauge group $SO(4,1)$ and $10$ generators ($6$ rotations and $4$ boosts).
Then via the current mechanism, $4$ of them become massive, and one is left with $6$ massless generators
($3$ rotations and $3$ boosts).
The reduced gauge group is $SO(3,1)$, corresponding to effective $4$-dimensions. The $5^{th}$ direction becomes redundant, and any dependence on its coordinate $x^4$ becomes trivial.
A gravity theory in $4$-dimensions would then necessarily be described by the Einstein-Hilbert action (with Newton's constant $G_N=1/M_{Pl}^{(4)2}$,
in terms of the only available mass/length scale),
the latter being unique for a spin $2$ field in that dimension \cite{feynman,deser,lovelock}.
This would be regardless of the starting Lagrangian $f(R)$ in $5$-dimensions in Eq.(\ref{act1}).
This mechanism provides an alternative to Kaluza-Klein
or brane-world compactification as a means of dimensional reduction,
and may have applications to string/M-theory for dimensional reduction from ten or eleven to the observed $4$-dimensions, one dimension
at a time, and at the symmetry
breaking scale in each dimension.
Observational evidence of the proposed mechanism
would be based on the predictions from the interaction terms in Eq.(\ref{act1}).

We end with the following interesting
observation: consider a
matter current coupling to the
{\it massive} spin connection via
$\mathcal{L}_{int}= - \lambda
\sum j_{ab}^\mu ~\omega_\mu^{ab}, $
where $\lambda$ is a dimensionless coupling constant (here the sum is over the massive spin-connections only).
At energies much lower than the four dimensional Planck mass  $M_\omega=M_{Pl}$,
the mass terms dominate, and
the effective Lagrangian  becomes
$
{\mathcal L}_{ eff}=
\lambda \sum 
\left[\frac{1}{2} M_\omega^2~\omega_{ab\mu}\omega^{ab\mu}
- \lambda j_{ab \mu} \omega_{ab}^{\mu} \right].
$
Varying this
with respect to the spin connection,  and substituting the stationary solution
$\omega_{ab \mu} = \frac{\lambda}{M_\omega^2} j_{ab \mu}~$ back
in the effective  Lagrangian, one obtains
\begin{eqnarray}
{\mathcal L}_{ eff}
&=&  -\sum
G_{N}
~j^{ab}_{ \mu} j_{ab}^\mu\hspace{2ex}\mbox{where}\hspace{2ex}
G_{N}
\equiv \left(\frac{\lambda}{\sqrt{2}M_{Pl}} \right)^2~.
\end{eqnarray}
Therefore at low energies, the dimensional
Newton's constant emerges from fundamental dimensionless coupling constant at high energies.
This is analogous to the emergence of
the effective dimensional Fermi constant $G_F$ at low energies in weak interactions, from  the
dimensionless $SU(2)$ coupling constant $g_2$,
$G_{F}= g^{2}_{2}/4\sqrt{2}M^{2}_{W}$, where  $M_W$ is  the mass of the $W$ boson
\cite{cottingham,chengli}.
Note however,  that such a simple interpretation cannot be used for
matter coupled to the long-ranged massless spin connections.

To summarize, in this paper we have proposed an alternative route to dimensional reduction of spacetime.
This happens via the Higgs mechanism, in which a Higgs field is coupled to the spin connections in gravity,
a suitable potential is introduced, and a symmetry breaking vacuum is chosen.
As a result, there is
spontaneous breakdown of a higher dimensional Lorentz group to one of lower dimensions, accompanied by the emergence of a massive Higgs field.
A lower dimensional Lorentz group implies an effective lower spacetime dimension seen for example by a gravity action.
This method of dimensional reduction
does not have the problems associated with the other methods such as Kaluza-Klein or brane world models.
It would be interesting to explore phenomenological consequences of this model and compare with observations to test if such a Higgs field actually exists in nature and if it can be a viable dark matter candidate.
It also remains to be seen if coupled to the pure gravity part of the action, this
mechanism provides a better ultraviolet behavior in four dimensions.
It may be noted that one can start with a ghost-free and renormalizable action (\ref{act1}) in higher dimensions,
for example the Lovelock action in dimensions greater than four \cite{lovelock2,zwiebach,boulware}.
Note that here we showed that a minimally coupled Higgs field achieved SSB, which was the goal of this work. However, it would be interesting to see what happens if one considered non-minimally coupled matter fields, e.g. dilaton or Brans-Dicke fields.
We hope to investigate these in the future.

\vspace{.2cm}
\noindent {\bf Acknowledgments}

\noindent
We thank P. Diaz, G. Kunstatter, R. B. Mann, J. Moffat, D. Smith, E. C. Vagenas and M. Walton for discussions. We thank the
Referees for their useful comments.
This work is supported by the Natural Sciences and Engineering Research Council of Canada.

\end{document}